\documentclass{optica-article}
\journal{opticajournal} % for journals or Optica Open
\articletype{Research Article}
\usepackage{lineno}
\definecolor{mycustompurple}{RGB}{46, 46, 177} % 定义自己的颜色
\usepackage{orcidlink}
\usepackage{verbatim}
\usepackage[normalem]{ulem}
\usepackage[commandnameprefix=always]{changes}
\setaddedmarkup{\textcolor{red}{#1}}
\arrayrulecolor{black}  % 设置表格线为黑色

\hypersetup{
	colorlinks=true,
	linkcolor=mycustompurple,
	citecolor=mycustompurple
}
\begin{document}
\title{A generative approach for lensless imaging in low-light conditions}

%~\orcidlink{0009-0008-0034-3755}
%~\orcidlink{0000-0002-6780-6100}
\author{ZIYANG LIU,\authormark{1}~\orcidlink{0009-0008-0034-3755} TIANJIAO ZENG,\authormark{2,*}~\orcidlink{0000-0002-6780-6100} XU ZHAN,\authormark{1}~\orcidlink{0000-0003-2816-9791}, XIAOLING ZHANG\authormark{1}~\orcidlink{0000-0003-2343-3055} and EDMUND Y. LAM\authormark{3}~\orcidlink{0000-0001-6268-950X}}

\address{\authormark{1}School of Information and Communication Engineering, University of Electronic Science and Technology of China, Chengdu, China\\
\authormark{2}School of Aeronautics and Astronautics, University of Electronic Science and Technology of China, Chengdu, China \\ 
\authormark{3}Department of Electrical and Electronic Engineering, The University of Hong Kong, Pokfulam, Hong Kong SAR, China \\ }

\email{\authormark{*}tzeng@uestc.edu.cn} %% email address is required; see note below about the corresponding author designation

% use {asbstract*} to suppress the copyright line. Copyright information will be added in production

\begin{abstract*} 
Lensless imaging offers a lightweight, compact alternative to traditional lens-based systems, ideal for exploration in space-constrained environments. However, the absence of a focusing lens and limited lighting in such environments often result in low-light conditions, 
where the measurements suffer from complex noise interference due to insufficient capture of photons.
This study presents a robust reconstruction method for high-quality imaging in low-light scenarios, employing two complementary perspectives: model-driven and data-driven. First, we apply a physic-model-driven perspective to reconstruct in the range space of the pseudo-inverse of the measurement model—as a first guidance to extract information in the noisy measurements. Then, we integrate a generative-model based perspective to suppress residual noises—as the second guidance to suppress noises in the initial noisy results. Specifically, a learnable Wiener filter-based module generates an initial, noisy reconstruction. Then, for fast and, more importantly, stable generation of the clear image from the noisy version, we implement a modified conditional generative diffusion module. This module converts the raw image into the latent wavelet domain for efficiency and uses modified bidirectional training processes for stabilization. Simulations and real-world experiments demonstrate substantial improvements in overall visual quality, advancing lensless imaging in challenging low-light environments\footnote{This manuscript has been accepted by Optics Express. © 2024 Optica Publishing Group. Users may use, reuse, and build upon the article, or use the article for text or data mining, so long as such uses are for non-commercial purposes and appropriate attribution is maintained. All other rights are reserved.}.
\end{abstract*}

%%%%%%%%%%%%%%%%%%%%%%%%%%  body  %%%%%%%%%%%%%%%%%%%%%%%%%%
\section{Introduction}
While lens technology has significantly propelled the progress of imaging science, its inherent physical constraints pose bottlenecks for further miniaturization, lightweight design, and cost reduction \cite{heide2013high, li2024lensless}. The contradiction between these physical constraints imposed by optical lenses on traditional imaging device sizes and the pursuit of miniaturization and thinness has sparked the emergence of lensless imaging technology. Lensless imaging follows the new evolution of ground-breaking computational imaging techniques. Through computational imaging—a tight integration of the sensing system and computation to form images of interest—we can access information that was otherwise not possible. This approach has shown promising performance across diverse areas such as holographic imaging \cite{zhang2024single}, phase recovery \cite{wang2024use, song2022iterative}, fluorescence microscopy \cite{huang2023enhancing, ge2023millisecond}, high dynamic range (HDR) imaging \cite{mai2022deep}, underwater imaging \cite{zhu2021full}, etc.

Lensless imaging utilizes simple and inexpensive optical encoders to replace costly and complex lens assemblies, leveraging computational imaging techniques to reconstruct clear images from collected measurements \cite{asif2016flatcam, boominathan2020phlatcam, boominathan2021recent, liu2023autofocusing, goswami2023assessment}. In lensless imaging, reconstruction is crucial due to the significant difference between measured data and the original scene. Most techniques use regularization-based methods to solve underdetermined linear equations, optimizing fidelity and data prior terms. Simple cases may use Tikhonov regularization for closed-form solutions, while complex scenarios require iterative algorithms like the iterative shrinkage-thresholding algorithm (ISTA) or the alternating direction method of multipliers (ADMM), offering better quality but with higher computational costs and manual parameter tuning.

Despite advancements, traditional model-based methods often fall short due to imprecise modeling of data priors and limitations in handling real-world complexities. Deep learning has introduced neural networks as powerful inversion operators, directly mapping raw measurements to imaging scenes through data-driven learning \cite{zhang2023dual}. For instance, Pan et al. \cite{pan2021incoherent} developed a transformer-based end-to-end reconstruction network. However, these methods often overlook the forward physical model, leading to image artifacts and loss of fine details. To bridge this gap, hybrid methods combine traditional optimization with deep learning. Monakhova et al. \cite{monakhova2019learned} introduced Le-ADMM-U, incorporating a neural network into an unrolled ADMM optimization loop, improving reconstruction by learning from data while maintaining optimization principles. Similarly, Khan et al. \cite{khan2020flatnet} proposed FlatNet, which refines a learnable Tikhonov-based reconstruction through a GAN with perceptual loss, enhancing image quality. One key challenge in hybrid methods is model mismatch—the difference between the assumed forward model and the actual system—which can degrade image quality. In our previous work \cite{zeng2021robust}, we quantified error accumulation from model mismatch and proposed a multi-stage information loss compensation method to improve reconstruction accuracy and stability. 
%Following our work, Kingshott \cite{kingshott2022unrolled} improved upon ADMM by implementing a learned optimization scheme through an unrolled primal-dual reconstruction method, which reduced model error. Li \cite{li2023mwdns} introduced a multi-scale Wiener deconvolution approach to recover lost information across different stages, while Qian \cite{qian2024robust} integrated a deep image denoising module into the iterative reconstruction process to minimize model error noise. More recently, Cai \cite{cai2024phocolens} developed a spatially-variable and learnable deconvolution method combined with a generative model for refinement.
Following our work, Kingshott \cite{kingshott2022unrolled} introduced a learned optimization scheme through an unrolled primal-dual reconstruction method to reduce model error. Li \cite{li2023mwdns} introduced a multi-scale Wiener deconvolution approach to recover lost information. Qian \cite{qian2024robust} integrated a deep denoising module into the iterative reconstruction process to minimize the model error. More recently, Cai \cite{cai2024phocolens} combined a spatially-variable learnable deconvolution method with a generative model for refinement reconstruction.

\subsection{Motivation and Aim}

Despite significant advances in reconstruction techniques, the performance of lensless imaging systems under low-light conditions remains an underexplored challenge. Without a focusing lens, these systems suffer from significant signal attenuation as light disperses through the mask, leading to reduced signal-to-noise ratios (SNR). This issue is further exacerbated by the small size of sensors, making high-quality imaging in resource-constrained or low-light environments particularly difficult. Most lensless imaging methods, such as those described in \cite{kingshott2022unrolled, cai2024phocolens, chakravarthula2023thin, bezzam2024let}, adopt a two-stage network design. The first stage incorporates a forward physical model to recover low-frequency image content, followed by a post-processing network (e.g., a denoiser or generative model) to refine and enhance the image. While these approaches yield promising results under normal lighting conditions, their performance degrades significantly in low-light scenarios due to the following limitations:
\begin{itemize}
    \item Noise Characteristics: In low-light conditions, the measurements are heavily influenced by complex noise patterns, which differ from those in normal lighting. Current two-stage methods often employ a denoiser or generative model in the second stage, but these are not optimized for low-light noise characteristics, leading to suboptimal results.
    \item Brightness Insufficiency: Low photon counts cause severely underexposed lensless measurements, posing challenges for existing network architectures in restoring brightness while maintaining fine image details and textures, often leading to unstable results and degraded reconstruction quality.
\end{itemize}

% Noise Characteristics: In low-light conditions, the measurements are heavily influenced by complex noise patterns, which differ from those in normal lighting. Current two-stage methods often employ a denoiser or generative model in the second stage, but these are not optimized for low-light noise characteristics, leading to suboptimal results.

% Brightness Insufficiency: Low photon counts in low-light environments result in severely underexposed measurements. Existing network architectures struggle to adequately restore brightness while preserving fine image details and textures, resulting in a trade-off that compromises reconstruction quality. Generative models, in particular, tend to produce unstable results under such conditions.

% These limitations motivate the approach proposed in this work, which targets the specific noise patterns and brightness insufficiency inherent to low-light conditions. Our method aims to provide a robust reconstruction framework that balances brightness restoration with detail preservation under noise interferences.

Therefore, this study aims to develop a robust reconstruction framework specifically designed for low-light lensless imaging, balancing brightness restoration, noise suppression, and detail preservation by leveraging the strengths of both physics-driven and generative models.

\subsection{Basic Idea and Contributions}
To enable lensless imaging in low-light conditions, our work builds upon two fundamental aspects: theoretical foundations and algorithmic methodologies. 

\begin{itemize}
    \item On the theoretical front, we present the first comprehensive analysis of noise characteristics inherent in low-light lensless imaging, and propose a theoretical model that serves as a foundation for designing reconstruction methods and generating simulation data tailored for network training. 
    \item Algorithmically, we propose a novel multi-step diffusion model explicitly conditioned on low-light illumination and intricate noise components within a two-stage reconstruction framework. Unlike methods for well-illuminated conditions which overlooks the complexity of photon-limited noise, we leverage wavelet-domain decomposition to separate brightness and noise in the latent space, serving as conditions  to target these issues directly, and employ multi-step diffusion process for superior noise suppression compared to one-step generative models. A bidirectional training strategy further ensures stability and robustness under challenging low-light scenarios.
\end{itemize}

% On the theoretical front, we thoroughly analyze the noise characteristics inherent in low-light lensless imaging process, proposing a theoretical model that serves as a foundation for designing reconstruction methods and generating simulation data tailored for neural network training. 

% Algorithmically, we design a generative model within a two-stage reconstruction framework to address challenges posed by complex noise patterns and insufficient brightness. Departing from current one-step generative approaches, we adopt a multi-step diffusion model, achieving superior noise suppression. Unlike the current diffusion models used for standard image generation in normal lighting conditions, our diffusion model is specifically conditioned on low-light illumination and intricate noise components. This makes it optimized for reconstructing images with both low brightness and noise. Additionally, we implement a bidirectional training strategy to stabilize the reconstruction process.

Specifically, we first analyze the forward measurement process of lensless imaging, examining each phase of data transition in detail based on the camera's characteristics in low-light conditions. This analysis establishes a model that accounts for two key features of lensless imaging results: complex noise patterns and insufficient brightness. This also provides us with the tools needed for subsequent dataset construction for neural network training.

Secondly, we follow a two-stage framework, leveraging the forward measurement model as a strong prior to guide the initial reconstruction. This allows us to obtain partial information of the imaging scene in the range space of its adjoint pseudo-inverse. The transition from measurement space to image range space more prominently reveals the two low-light features mentioned above. 

Third, we employ a diffusion model to refine the initial result, addressing the two low-light features through a conditional approach. We incorporate these features into the diffusion model's generation process. Specifically, we decompose the initial result through wavelet transforms to separate brightness and noise information in the latent space, using these as conditions for the generative model. For the nullspace refinement, we refine the remaining texture information separately through a depth-separable convolutional neural network. This separation also allows the generation process to occur in a smaller latent space, enabling memory-efficient training and testing. Additionally, to address the increased instability of generation in underdetermined low-light conditions, we implement a bidirectional training strategy—incorporating both generation and diffusion processes—to stabilize the final imaging result.

To thoroughly evaluate the effectiveness of our proposed method, we conducted a series of both simulated and real-world experiments using a self-built, lensless camera within a carefully controlled lighting environment. For a comprehensive comparison, we employed both traditional non-learning-based approaches and cutting-edge learning-based models. The results are telling: conventional methods experience significant performance degradation, particularly under photon-limited conditions, where some even fail entirely. In contrast, our newly proposed method not only holds up but shines—quite literally. It demonstrates remarkable improvements in image brightness, superior noise reduction, and a clear enhancement in overall image quality.

% \XX{The contributions of this work can be summarized as follows:}
% \XX{\begin{itemize}
%     \item An in-depth theoretical analysis of noise characteristics inherent in low-light lensless imaging, providing insights into dataset generation and network design.
%     \item A two-stage reconstruction framework that integrates physics-based priors with a conditional diffusion model. The framework utilizes wavelet-domain decomposition to separate brightness and texture information for targeted refinement and employs a bidirectional training strategy to ensure stable and robust reconstructions under low-light conditions.
%     \item Comprehensive evaluations using both simulated and real-world datasets, demonstrating significant improvements in overall image quality compared to existing methods.
% \end{itemize}}

\section{Problem analysis}\label{s0}

\begin{figure}[t]
\centering\includegraphics[width=\textwidth]{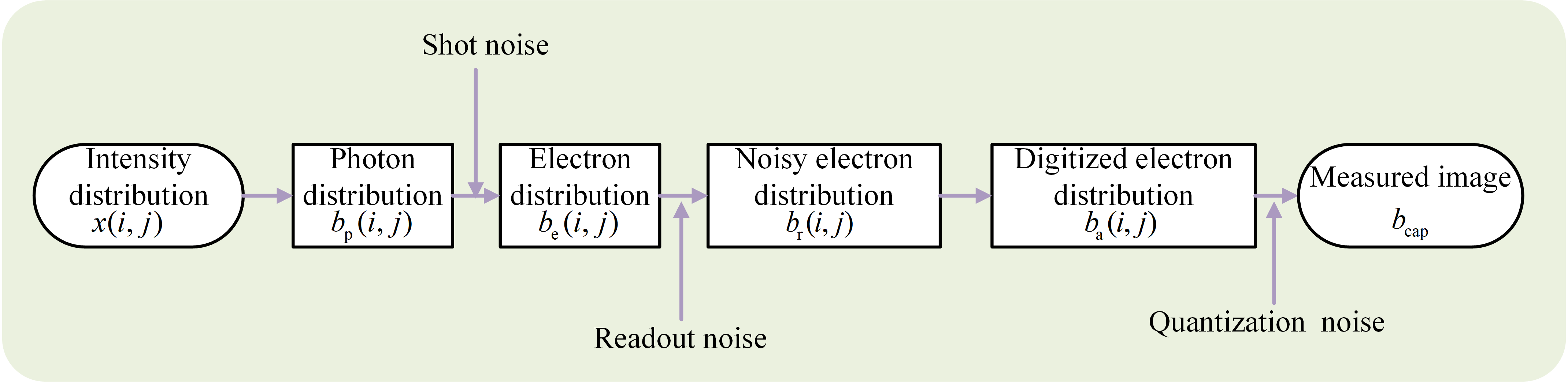}
\caption{The measurement process in low-light conditions, illustrating the mixture of multiple noise types.}
\label{f1}
\end{figure}

In this section, we analyze the impact of low-light conditions on lensless imaging from the perspective of the measurement process \cite{goswami2024robustness}, as illustrated in Fig.~\ref{f1}. Let's consider an intensity distribution to be measured, denoted as $x(i,j)$. This distribution undergoes a linear conversion to a photon distribution:
\begin{equation}
{b_p}(i,j) = K \times x(i,j)
\end{equation}
where $K$ represents the photon conversion efficiency. In low-light conditions, photon conversion efficiency decreases, resulting in a significant reduction in photon numbers. The process of photons reaching and being captured by the sensor follows a random Poisson process, introducing Poisson noise (also known as shot noise). This noise is amplified due to the reduced photon count. The captured photons are then linearly converted to electrons:
\begin{equation}
{b_e}(i,j) = \eta \times {\rm{Poisson}}({b_p}(i,j))
\end{equation}
where $\eta$ denotes the quantum efficiency, and ${\rm{Poisson}}\left( \lambda \right)$ is an operator that samples a Poisson random variable with mean $\lambda$.
This process also introduces additive Gaussian noise, known as readout noise. The resulting electron distribution becomes:
\begin{equation}
{b_r}(i,j) = {b_e}(i,j) + {n_r}
\end{equation}
where $n_r \sim \mathcal{N}(0, \sigma^2)$ and $\sigma$ represents the standard deviation of the readout noise. Subsequently, this electron distribution is digitized with a certain bias and quantized into the measured image:
\begin{equation}
{b_a}(i,j) = d \times {b_r}(i,j) + {b_l}
\end{equation}
where $d$ denotes the analogue-to-digital conversion operation, and ${b_l}$ is the bias amount. The digital image is then quantized for storage, and the final captured image can be expressed as:
\begin{equation}
{b_{cap}}(i,j) = {\rm{Quantize}}({b_a}(i,j))
\end{equation}
where Quantize(·) denotes the quantization operation, which introduces additional uniform noise into the stored digital image.

Throughout this process, multiple types of noise accumulate. In low-light conditions, the severe lack of photon capture and significant amplification of Poisson noise serve as initial sources that compound subsequent noise effects. This combination weakens the measured image quality, ultimately complicating the process of reconstructing and recovering lensless measurements in low-light conditions.

\section{Proposed Method}\label{s1}
\begin{figure}[t]
	\centering\includegraphics[width=\textwidth]{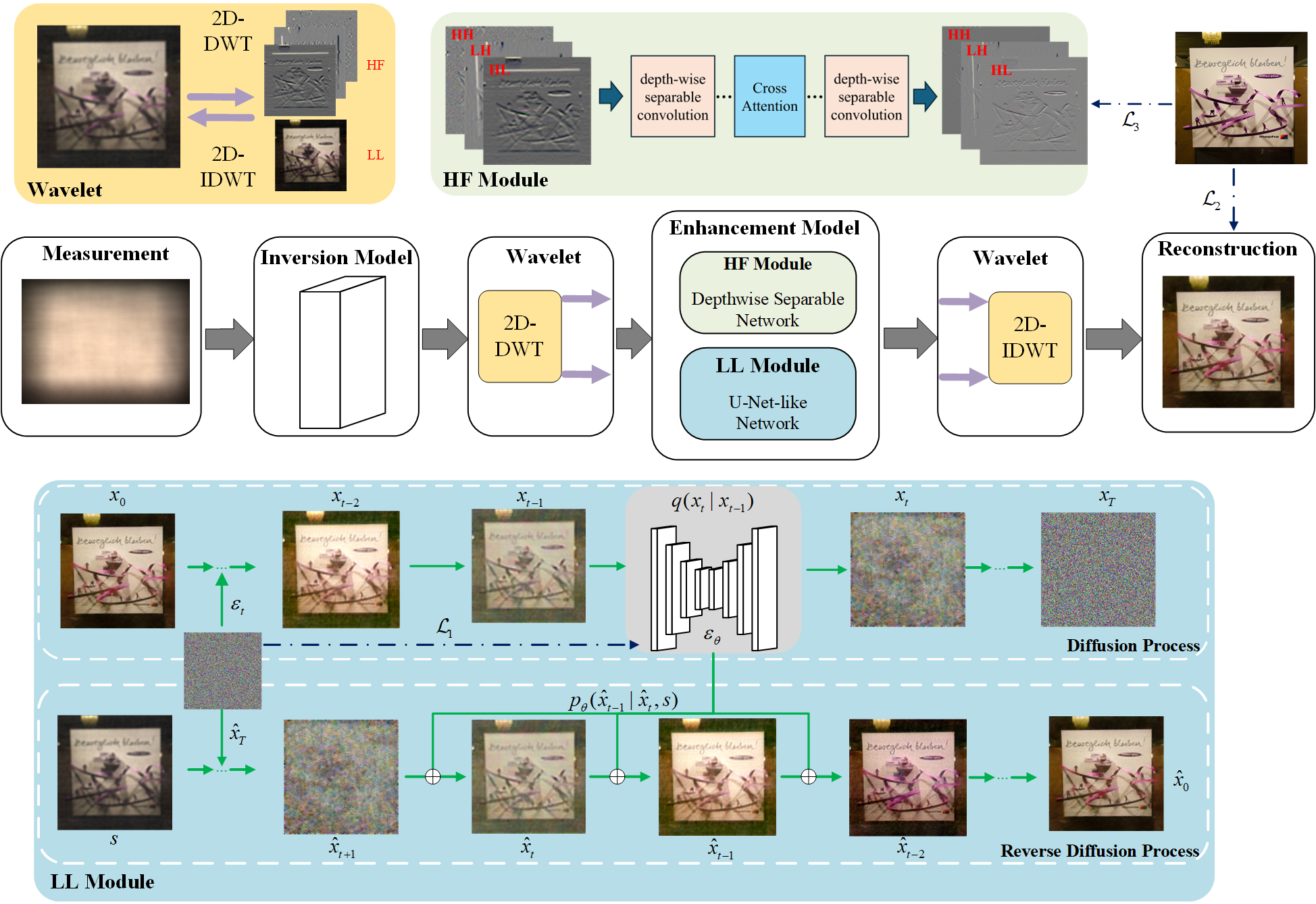}
	\caption{Overview of the reconstruction pipeline for the proposed framework.}
	\label{f2}
\end{figure}
Analysis from the previous section reveals that the dominant challenge is noise interference, which necessitates a reconstruction method resistant to such interference. In this context, we introduce our proposed method. We begin with a general explanation of our methodology and then delve into the specifics in the following subsections.

Intuitively, the severe noise interference in the measured image degrades the information we can extract directly, making it challenging to use a network to map the relationship from the noisy image to the scene. Therefore, we employ a "closer-to-closer" strategy. We fully utilize the physics model as a prior to guide an initial noisy reconstruction, then progressively refine it to achieve a clear reconstruction through data-driven mapping. Fig.~\ref{f2} illustrates the entire framework of our method. We will now break it down into more technical details.

\subsection{Fisrt Stage}\label{s2}
In the first stage, we rely on the forward measurement model of lensless imaging as follows:
\begin{equation}
{\bf{b}} = {\bf{H}}{\bf{x}}
\label{eq1}
\end{equation}
where ${\bf{b}}$ denotes the measurements collected by the sensor, ${\bf{H}}$ represents the forward measurement process of the system (the convolutional matrix of the system's point-spread function (PSF) obtained through practical calibrations), and ${\bf{x}}$ denotes the measured scene.

From this linear equation, we can see that partial information of the scene ${\bf{x}}$ lies in the range space of the adjoint operator of the forward measurement process:
\begin{equation}
{\bf{H^{+}b}} = {\bf{H^{+}H}}{\bf{x}} = \bf{x^{+}}
\label{eq1_1}
\end{equation}
where $\bf{H^{+}}$ is the adjoint operator of the forward measurement process. Considering the orthogonal decomposition $\bf{x} = (\bf{H^{+}H})\bf{x} + (\bf{I} - (\bf{(H^{+}H)})\bf{x} = \bf{x^{+} + x^{-}}$, $\bf{x^{+}}$ represents the range component of ${\bf{x}}$.

For fast computation, we turn to the direct inverse in the frequency domain, known as Wiener filtering:
\begin{equation}
{\bf{\hat x}} = {{\cal F}^{ - 1}}\left\{ {{\cal F}({\bf{b}}) \odot \left( {\frac{{{\cal F}{{({\bf{h}})}^*}}}{{({\bf{\lambda }} + {\cal F}{{({\bf{h}})}^{\bf{2}}})}}} \right)} \right\}
\end{equation}
where $\lambda$ is a noise-related factor (fixed in the experiment), ${\cal F}$ and ${{\cal F}^{ - 1}}$ represent the Fourier transform and its inverse, respectively, and $\bf{h}$ denotes the PSF. Here, it is initialized with the calibrated one but is set as learnable.

This direct inverse provides partial information of ${\bf{x}}$ embedded in the range space. However, as the measurement $\bf{b}$ is highly noisy, the obtained information of ${\bf{x}}$ is still affected by noise, not entirely accurate, and the part of information $\bf{x}-\bf{x^{+}}$ is still missing. Consequently, these initial results suffer from issues such as amplified noise, extremely low brightness, and poor readability, as seen in the experimental results. To address this, in the next stage, we adopt a diffusion generative model to suppress the noise and progressively generate the missing information.

\subsection{Second Stage}\label{s3}
\subsubsection{Conditional Diffusion Model}
In the second stage, we implement a sophisticated data-driven diffusion generative model. This model's core principle is to gradually generate the distribution of the target image $\bf{x_0}$ from noise, following a meticulously designed multiple-step Markov chain. The term "diffusion" aptly describes the inverse of the generation process, as noise is systematically introduced into the clear image—effectively diffusing it.

The relationship between adjacent images $\bf{x_t}$ and $\bf{x_{t-1}}$ in this diffusion process can be mathematically expressed as follows \cite{ho2020denoising}:
\begin{equation}
q({\bf{x_t}}|{\bf{x_{t - 1}}}) = {\cal N}(\bf{x_t};\sqrt {{\alpha_t}} {\bf{x_{t - 1}}},(1 - {\alpha _t}){\bf{I}})
\end{equation}
Here, $\alpha\bf{_t}$ represents predefined diffusion parameters. As the steps are sufficiently close, we can approximate the added noise as Gaussian. Through successive accumulation steps, the final diffused image converges to a normal distribution.

Conversely, in the generation process, we can relate these two images using the Bayesian theorem:
\begin{equation}
{p_\theta}({\bf{x}_{t - 1}}|{\bf{x_t}}) = {\cal N}({\bf{x_{t - 1}}};{\mu _t}(\bf{x_t},t),{\tilde \beta _t}{\bf{I}})
\label{eq2}
\end{equation}
The mean $\mu \bf{_t} $ and variance $\tilde \beta \bf{_t}$ in this equation are expressed as:
\begin{equation}
{\mu _t}(\bf{x_t},t) = \frac{1}{{\sqrt {1 - {{\bar \alpha }_t}} }}(\bf{x_t} - \frac{{{\beta _t}}}{{\sqrt {1 - {{\bar \alpha }_t}} }}{\varepsilon_\theta(\bf{x_t} , t)}), {\tilde \beta_t} = \frac{{1 - {{\bar \alpha }_{t - 1}}}}{{1 - {{\bar \alpha }_t}}}{\beta _t}
\end{equation}

In these equations, ${\beta \bf{_t}} = 1 - {\alpha \bf{_t}}$, ${\bar \alpha \bf{_t}} = \prod\nolimits_1^T {{\alpha \bf{_t}}}$, and $T$ denotes the total number of generation steps. The term ${\varepsilon _\theta }(\bf{x_t},t)$ denotes the noise added during the diffusion process and is the key variable that the generation module must learn to predict. For a deeper dive into the intricacies of the diffusion model, we refer readers to the work by Ho et al. \cite{ho2020denoising}.

The original diffusion model described above is intended for general image generation tasks, where a high-quality realistic image can be generated. However, for our task, we have one important condition to consider: the generated image must adhere to the measured image using the lensless forward measurement model. In other words, the generation process must be guided, becoming less random, to produce the result we require. In this context, we design two techniques: one for the generation process itself and another for the training process. 

To enhance the generation process, we introduce a conditional distribution approach, which is the LL Module in Fig~\ref{f2}. By incorporating the initial reconstruction result as an additional input to the network, we effectively condition the generation on the context of the imaging scene and the measurement process noise. This can be formalized as:
\begin{equation}
	{p_\theta}(\bf{x}_{t-1})|{\bf{x}_t},s) = {\cal N}({\bf{x}_{t - 1}};{\mu _\theta }({\bf{x}_t},s,t),{\rm{ }}{\tilde \beta _t}{\bf{I}})
\end{equation}
Here, $\mathbf{s}$ represents the initial reconstructed result from the first stage, encapsulating the low-light conditions. Consequently, the noise prediction task is reformulated as $\varepsilon_\theta(\mathbf{x}\bf{_t},\mathbf{s},\bf{t})$

Secondly, to mitigate potential instabilities during inference, we implement a comprehensive training regimen. This approach requires the network to execute both the forward diffusion process—where random Gaussian noise is systematically added to both the high-quality image and the conditioned initial reconstructed result under low-light guidance—and the reverse generation process. The latter involves continuous noise removal based on the neural network's learned priors. During the testing phase, only the reverse generation process is employed, wherein the initial reconstructed result and a randomly Gaussian-distributed image undergo progressive denoising and enhancement, leveraging the network's learned priors to yield the desired high-quality, realistic image.

Specifically, we first preprocess the preliminarily reconstructed image from the first stage. The Wavelet Transform can significantly reduce the spatial dimension of images without losing information. We utilize the Haar Discrete Wavelet Transform (DWT) \cite{jiang2023low} to transform the preliminarily reconstructed image into a higher-dimensional wavelet domain. By decomposing the image, we obtain four smaller sub-bands: the low-frequency component, and the high-frequency components in the horizontal, vertical, and diagonal directions. This transformation can be expressed as:
\begin{equation}
\begin{array}{l}
\left\{{\bf{LL},\bf{LH},\bf{HL},\bf{HH}} \right\} = {2{\rm{D-DWT}}} \left\{\bf{x}\right\}\\
\hat{\bf{x}} = 2{\rm{D-IDWT}}\left\{ {\hat {\bf{LL}},\hat {\bf{LH}},\hat {\bf{HL}},\hat {\bf{HH}}} \right\}
\end{array}
\end{equation}
where 2D-DWT and 2D-IDWT represent the 2D Discrete Wavelet Transform and the 2D Inverse Discrete Wavelet Transform, respectively. $\bf{x}$ denotes the input image, $\bf{LL}$ represents the low-frequency information, while $\bf{LH}$, $\bf{HL}$, and $\bf{HH}$ represent the high-frequency information in the vertical, horizontal, and diagonal directions, respectively. The hatted variables denote the corresponding reconstructed images. By applying the wavelet transform twice, we reduce the image resolution by a factor of four, lowering memory and computational demands while preserving key information for the diffusion model. This process decomposes the image into low- and high-frequency components. The low-frequency component retains global structural information, while the high-frequency component captures fine details. This separation allows the conditional diffusion model (LL module) to focus on low frequencies, enhancing brightness, reducing noise, and recovering basic contours. Meanwhile, the depthwise separable convolution network (HF module) targets high frequencies, enhancing textures and fine details.

\subsubsection{Processing structure}

In the second stage, the LL Module and HF Module are employed to further denoise and enhance the coarse reconstruction results from the first stage. Specifically, a wavelet transform is applied to decompose the initial reconstruction into low-frequency (LL) and high-frequency (HF, including HH, LH, and HL) components. The LL Module utilizes a conditional diffusion model to process the low-frequency sub-band $\bf{LL}$ extracted from the wavelet-transformed coarse result. By concatenating the low-frequency sub-band from the initial reconstruction with the corresponding sub-band from a normal-light reference image, the diffusion model is guided to generate a high-quality normal-light sub-band image from the noisy low-light input. Simultaneously, the HF Module leverages a depth-wise separable convolutional network to denoise and restore the high-frequency sub-bands. Through a cross-attention mechanism, the network enhances feature interactions among the high-frequency sub-bands (HH, LH, HL), ultimately producing optimized results. This two-module design ensures effective enhancement of both low-frequency and high-frequency information, significantly improving the overall quality of the reconstructed image.

We utilize a deep separable convolutional network within the HF Module, as illustrated in Fig~\ref{f2}, to restore fine details and high-frequency information extracted from the wavelet transform sub-bands (HH, LH, HL). This module is designed to enhance image clarity and texture by effectively processing and fusing high-frequency components. Initially, depth-wise separable convolution is employed to preliminarily extract features from the input sub-bands. This approach processes each channel independently, significantly reducing computational complexity while preserving essential details. The extracted features then interact through a cross-attention mechanism, which captures correlations and complementary information across different frequency components. This step facilitates more accurate feature fusion. Following feature fusion, the features undergo further refinement through additional depth-wise separable convolution layers, enhancing feature representation and improving network robustness. The processed sub-bands (HH, LH, HL) are then output as optimized high-frequency feature maps. By integrating efficient convolution operations with an attention mechanism, this design effectively extracts and fuses high-frequency information, improving the image's overall texture and detail quality.

\subsection{Loss Function}\label{s4}
The network first employs mean squared error (MSE) loss to constrain the forward diffusion process of the diffusion model, aiming to reduce the discrepancy between the predicted noise and the added noise, as shown in the following equation:
\begin{equation}
	{\cal L}_{1} = {{\rm{E}}_{\bf{t},{x_0},{\varepsilon \bf{_t}}}}[{\left\| {\varepsilon \bf{_t} - {\varepsilon _\theta }({x\bf{_t}},s,\bf{t})} \right\|^2}]
\end{equation}

Given the instability of the reverse diffusion process in the proposed network, a combination of mean absolute error (MAE) loss, Structural Similarity Index Measure (SSIM) loss \cite{wang2004image}, and learned perceptual image patch similarity (LPIPS) \cite{zhang2018unreasonable} is utilized to constrain the reverse diffusion process, which is also the network reconstruction enhancement process, as formulated below:

\begin{equation}
			{\cal L}_{2} = \, {\lambda_1}{\left\| {\hat{\bf{x}} - \bf{x}} \right\|_1} + {\lambda_2}{\rm{SSIM}(\hat{\bf{x}}, \bf{x})} + {\lambda_3}\left\{ \left\| {{f_2}(\hat{\bf{x}}) - {f_2}(\bf{x})} \right\|^2 
			+ \left\| {{f_4}(\hat{\bf{x}}) - {f_4}(\bf{x})} \right\|^2 \right\}
\end{equation}
where $\hat{\bf{x}}$ and $\bf{x}$ denote the reconstructed enhanced image and the ground truth respectively, $f_2$ and $f_4$ denote the second convolutional layer and the fourth convolutional layer of the pre-trained network, and ${\lambda _1},{\lambda _2},{\lambda _3}$  represents the weight of each loss term.

Furthermore, a combination of MSE loss and Total Variation (TV) loss is employed to constrain the reconstruction of high-frequency information in the image, as shown in the equation below:
\begin{equation}
		{\cal L}_{3} = {\lambda _4}{\left\| \hat{\rm{HF}} - {\rm{HF}} \right\|^2} + {\lambda _5}{\rm{TV}}(\hat{\rm{HF}}, {\rm{HF}})
\end{equation}
where $\hat{\rm{HF}}$ and $\rm{HF} $ represent reconstruction of the enhanced high-frequency component and ground truth of high-frequency component, and ${\lambda _4},{\lambda _5}$ represents the weight of each loss term.

In summary, ${{\cal L}_{1}}$ ensures accurate noise prediction during the forward diffusion process in our conditional diffusion model. ${{\cal L}_{2}}$ facilitates high-quality image generation during the reverse diffusion process. Finally, ${{\cal L}_{3}}$ emphasizes the reconstruction of the enhanced high-frequency components. As marked in Fig~\ref{f2}, the total loss function of the proposed network is:
\begin{equation}
	{{\cal L}_{{\rm{all}}}} = {{\cal L}_{\rm{1}}} + {{\cal L}_{\rm{2}}} + {{\cal L}_{\rm{3}}}
\end{equation}

\section{Experiment and Results}
\subsection{Dataset}
Due to the lack of publicly available low-light lensless imaging datasets, we simulated measurements using an established lensless imaging model and actual measured PSF. We used the LOLv2 dataset\cite{yang2021sparse}, selecting 1000 pairs of synthetic low-light and normal-light images. These were processed through our lensless imaging model to create a low-light lensless dataset, with 900 pairs for training and 100 for testing.

To validate our method in real-world scenarios, we developed a lensless camera. We projected images onto an LCD screen and captured measurements by adjusting the camera's acquisition time and exposure via a Raspberry Pi. This approach aligns with actual lensless camera imaging and facilitates labeled dataset collection. We used the "Synthetic" subset of LOLv2, processing the normal-light images for projection and pairing them with captured low-light lensless measurements.

\subsection{Impletmentation Details}
The prototype of the lensless camera used in this experiment employs a camera equipped with a IMX219 CMOS sensor, featuring a pixel size of 1.12µm. The dimensions of all ground truth images are adjusted to 384×384, equivalent to the calibrated camera's field of view, ensuring consistency in size between the input images and ground truth images for the network. We directly utilize Bayer measurements, divided into four channels (R, Gr, B, Gb), as the input for raw imaging, utilizing the full size of 2028×1520×4. 

The implementation of our experiments is accomplished within the PyTorch framework. The $\lambda$ parameter in the Wiener filter controls noise suppression, initially set to 50000(as in \cite{khan2020flatnet}) and increased to 80000 for noisier scenes. However, noise reduction is mainly handled by the second-stage diffusion model, which has a greater impact on the final image quality. After the first stage, the region of interest is cropped to 384×384×3. For training, images are randomly cropped into 256×256×3 patches. During testing, the reconstructed image is kept at 384×384×3 without cropping. The Adam optimizer is utilized to train the network for 500 epochs with an initial learning rate of $10{^{-4}}$, decayed by 0.8 every 100 epochs. No weight decay is applied. Exponential Moving Average (EMA) is implemented on model parameters at a rate of 0.9999 to ensure a more stable training process. The dropout value for the resnet blocks within the model is set to 0.3. During the training phase, the diffusion step size T is set to 200, the implicit sampling step size is set to 10, and the batch size is 22. The entire experimental process is executed on a Windows system equipped with 32GB of RAM and two NVIDIA RTX 3090 GPUs.

\subsection{Quantitative Metrics}
In addition to qualitative evaluations based on human visual perception, this paper also selects a range of quantitative metrics to effectively assess the experimental results. Apart from the classic MSE to measure the degree of image quality loss, Peak Signal-to-Noise Ratio (PSNR) to reflect the fidelity of image signals, and SSIM to evaluate the similarity of image structures, we have additionally incorporated LPIPS , an index that aligns more closely with human visual perception, as a metric. Unlike traditional error-based evaluation metrics, LPIPS is an image quality assessment metric based on a trained neural network model. It aims to capture differences in human perception by comparing the local perceptual features of two images. These features are obtained by training a deep convolutional neural network on a large dataset of image pairs, where the network learns to map image content into a low dimensional space where images that are perceptually similar to humans have smaller distances. LPIPS considers not only pixel-level differences but also perceptual differences, making it better at predicting human subjective perception. This comprehensive approach ensures an objective and accurate evaluation of the experimental outcomes, while better capturing the nuances of human visual experience.

\subsection{Simulated Reconstruction }
In the simulation experiments, a point light source was placed 320 mm in front of the random binary mask and 150 mm in height, and the PSF was acquired using the lensless camera constructed in this paper. The random binary mask is 10 mm away from the CMOS sensor. The output resolution of our sensor is 2028 × 1520 with a pixel pitch of 0.014 mm. Based on the forward imaging model of the lensless camera in Section \ref{s2}, a simulated dataset was obtained using the captured PSF and existing low-light images.

First, the proposed reconstruction enhancement method is trained and evaluated using the simulated training and test sets. To comprehensively demonstrate the effectiveness of the low-light lens-free reconstruction enhancement method introduced in this paper, we have deliberately selected several well-established methods that perform well under normal lighting conditions for comparison. These methods include ADMM \cite{bezzam2022lenslesspicam} with 100 iterations, the purely data-driven U-Net \cite{monakhova2019learned}, FlatNet \cite{khan2020flatnet}, which combines generative adversarial networks and perceptual losses, MWDN \cite{li2023mwdns} with multi-scale deconvolution, and DeepLIR \cite{poudel2024deeplir}, a two-stage network integrated with an attention mechanism. Unlike previous studies, however, this experiment applies these methods to low-light conditions to assess their actual performance.

\begin{figure}[htbp]
	\centering\includegraphics[width=\textwidth]{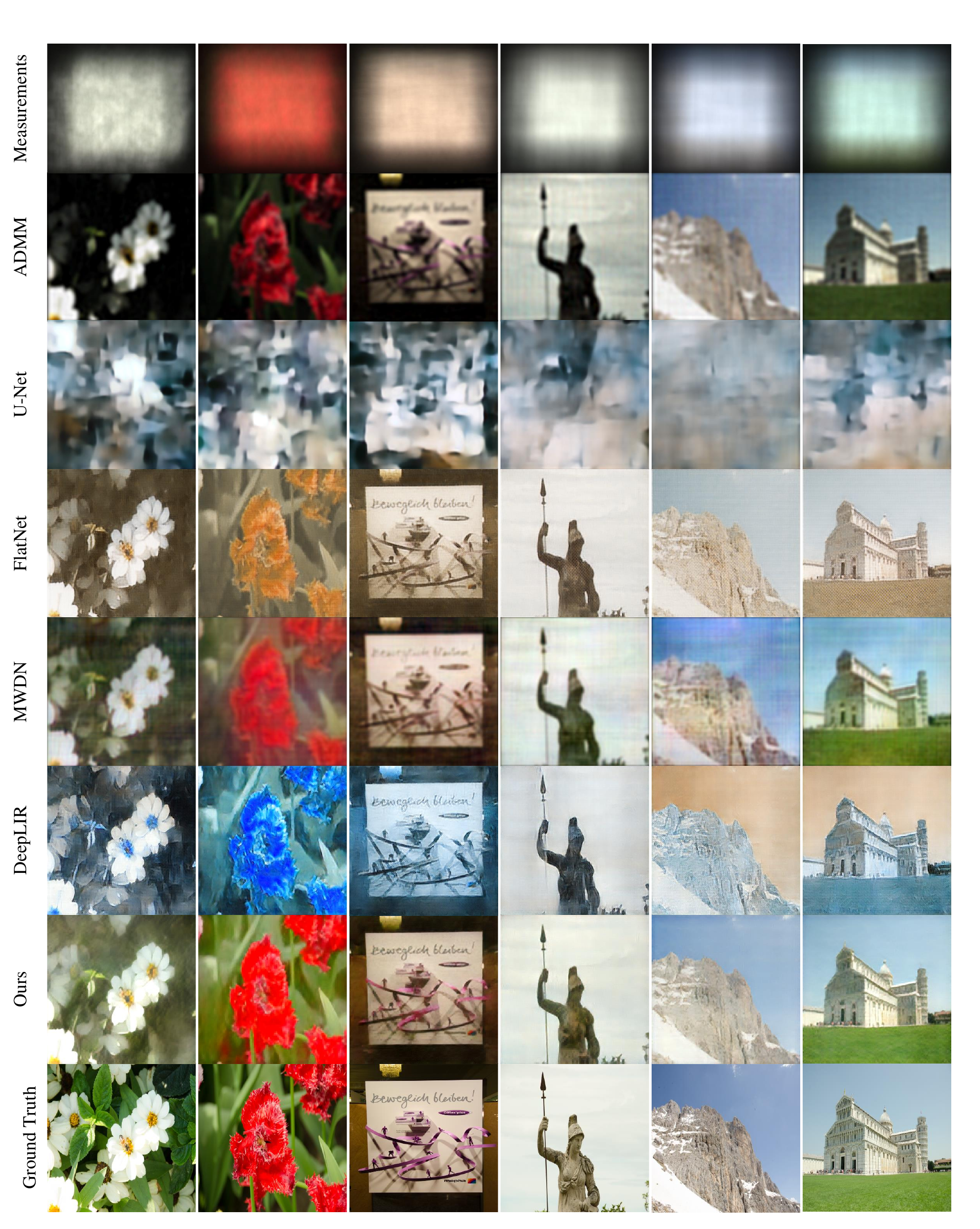}
	\caption{The test set results for the simulated dataset, from top to bottom, are the Measurements, ADMM, U-Net, FlatNet,  MWDN, DeepLIR, Ours, and Ground Truth.}
	\label{f3}
\end{figure}

Fig.~\ref{f3} shows the reconstructed images under low-light conditions using different methods, along with the original input and ground truth images. Visual comparison reveals that although these classical methods perform well under normal lighting, their reconstruction results are significantly degraded under low-light conditions, exhibiting noticeable blur, distortion, and color shift. In contrast, the method proposed in this paper maintains high reconstruction quality even under low-light conditions, with clear image details and accurate color restoration, demonstrating its unique advantages in low-light, lens-free reconstruction and enhancement.

Specifically, compared to traditional optimization methods, generic data-driven networks, physics-driven networks, and data-driven two-stage networks, the images reconstructed by the proposed model exhibit superior visual quality across all samples. While ADMM can recover the basic contours from the raw measurements, it fails to effectively enhance the image brightness, resulting in overall dark reconstruction with hidden details. As a purely data-driven method, U-Net is unable to generate accurate scene images, indicating its limited capability when working with small datasets. FlatNet, which combines generative adversarial networks and perceptual losses, improves reconstruction quality but still struggles with restoring fine details and color accuracy. MWDN achieves better results but still falls short in recovering precise details and brightness. Similar to FlatNet, DeepLIR suffers from significant color distortion. This highlights the increased complexity of data characteristics under low-light conditions. The significant brightness disparity creates a need for brightness enhancement, which causes the original denoising network to lose focus on accurate color and detail restoration. In contrast, the model proposed in this paper shows clear advantages under low-light conditions, with reconstructed images that are closer to the ground truth and richer in both color and detail, thanks to the model’s careful consideration of the unique characteristics of low-light data and its targeted optimization strategies during training.

\begin{table}[htbp]
	\centering
	\caption{\bf The average MSE, LPIPS, PSNR and SSIM of the proposed method and several other methods on the simulation test set.}
	\begin{tabular}{ccccc}
		\hline
		Method  & MSE & LPIPS & PSNR(in dB) & SSIM\\
		\hline
		ADMM    &  0.1009& 0.3666 & 11.00 & 0.3283\\
            U-Net    &  - & - & - & -\\
		FlatNet & 0.0259 & 0.2099 & 17.05 & 0.4647\\
            MWDN    &  0.0190 & 0.2646 & 17.7218 & 0.6115\\
            DeepLIR    & 0.0636 & 0.2720 & 13.6968 & 0.4463\\
		Ours & 0.0166& 0.1605& 18.83& 0.5719 \\
		\hline
	\end{tabular}
	\label{tab:0s}
\end{table}

To further quantify the analysis, Table~\ref{tab:0s} presents the average MSE, LPIPS, PSNR, and SSIM of each algorithm on the simulated test dataset. The traditional ADMM method shows poor performance across all metrics due to the high noise and low brightness in the reconstructed images. FlatNet and DeepLIR, as two-stage networks, are able to perform image reconstruction but still struggle with color and detail restoration, leading to suboptimal performance in all metrics. MWDN, by performing reconstruction in a multi-scale space, achieves relatively better results, particularly in SSIM. However, the proposed method combines the physical model of lens-free reconstruction with low-light priors, allowing it to outperform the others across all metrics, demonstrating superior reconstruction quality.

These results not only validate the effectiveness of the proposed method but also highlight the limitations of existing imaging techniques under low-light conditions, further emphasizing the need for specific optimizations and designs for low-light environments.

The lensless camera noise model in Section \ref{s0} allows for a more accurate simulation of the complex noise characteristics generated during actual CMOS imaging. The model contains a full set of noise components, mainly read noise, Poisson noise and quantisation noise. In order to further validate the robustness and effectiveness of the proposed method, we inject noise into the original measurements in the simulated dataset according to the above noise model. This approach ensures that the features of the simulated dataset are very similar to those of the real-world measurements, which improves the reliability and credibility of the experimental results. Table \ref{tab:1s} details the values of the key parameters involved in the implementation.
\begin{table}[htbp]
	\centering
	\caption{\bf The simulation parameter values of the camera noise added to the simulation data set.}
	\begin{tabular}{lc}
		\hline
		Parameters  & Values \\
		\hline
		The maximum light intensity of the camera &  1000\\
		The quantum efficiency of the camera  & 0.7  \\
		The standard deviation of read noise & 2.63 \\
		The Analog to Digital Unit (ADU) of the camera & 0.23 \\
		The baseline ADU of the camera &  4.48\\
		The number of bits of the camera & 8\\
		\hline
	\end{tabular}
	\label{tab:1s}
\end{table}

\begin{figure}[t]
	\centering\includegraphics[width=\textwidth]{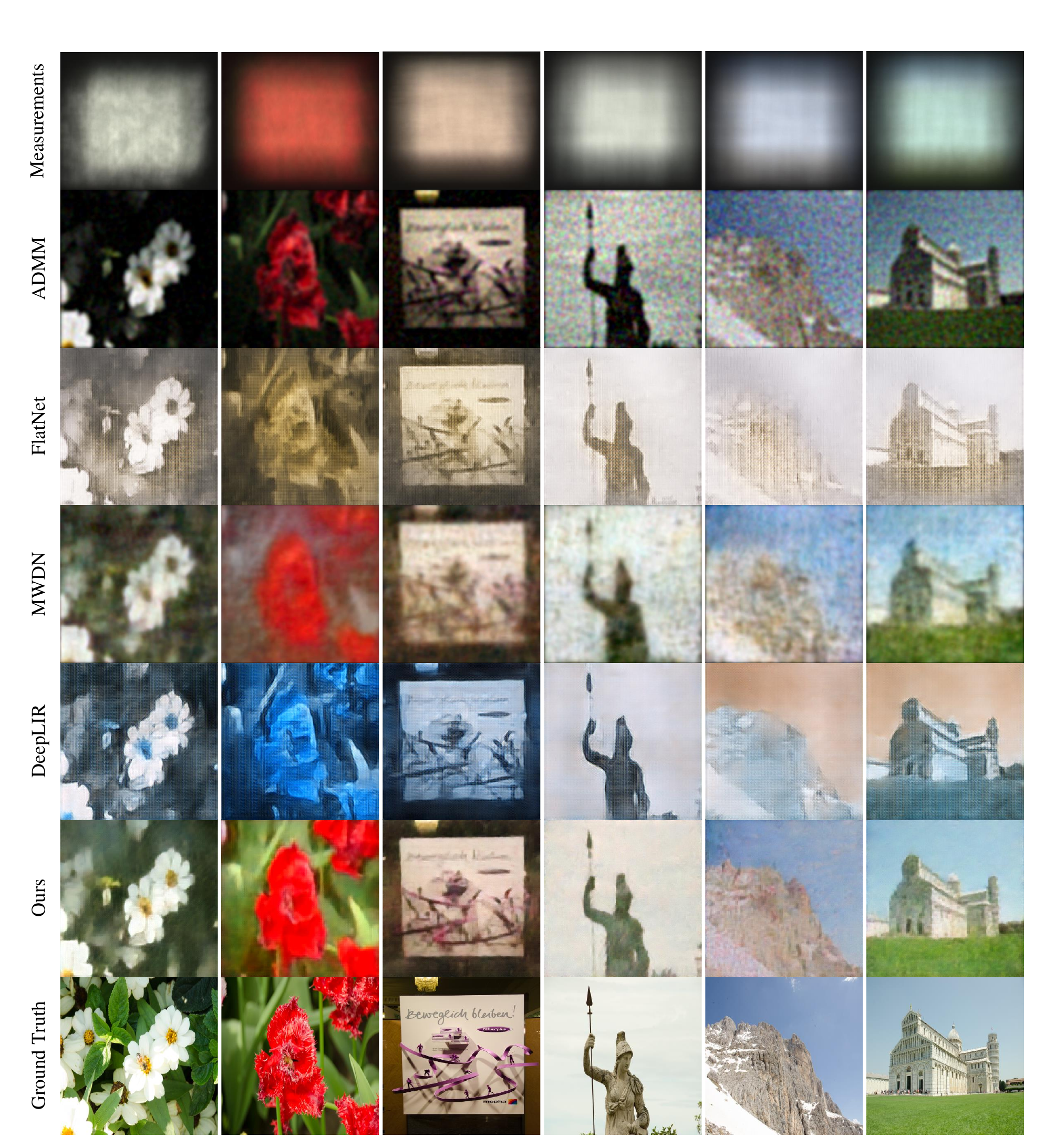}
	\caption{Reconstruct and enhance results in a simulated dataset with camera noise, from top to bottom, are the Measurements, ADMM, FlatNet, MWDN, DeepLIR, Ours, and Ground Truth.}
	\label{f4}
\end{figure}

 Fig.~\ref{f4} presents the reconstruction examples of various methods on the simulated test dataset with added camera noise. As shown in the figure, ADMM successfully recovers most of the image structure but is heavily contaminated by complex noise, which obscures fine details and does not improve image brightness. While FlatNet and DeepLIR are effective at removing most of the noise and enhancing brightness, they suffer from significant loss of detail and color information. MWDN achieves basic reconstruction and ensures color recovery, but still falls short in terms of noise suppression and fine detail restoration. In contrast, the proposed method not only reduces noise effectively but also significantly enhances image brightness, resulting in visually acceptable reconstruction and enhancement. This demonstrates that our method is highly robust to noise.

\begin{table}[htbp]
	\centering
	\caption{\bf The average MSE, LPIPS, PSNR and SSIM of the proposed method and several other methods on the simulation test set.}
	\begin{tabular}{ccccc}
		\hline
		Method  & MSE & LPIPS & PSNR(in dB) & SSIM\\
		\hline
		ADMM    & 0.1048& 0.4675 & 10.70 & 0.2990\\
		FlatNet & 0.0482 & 0.2321 & 14.02 & 0.3350\\
             MWDN   &  0.0249 & 0.3003 & 16.4291 & 0.5275\\
            DeepLIR    &  0.0579 & 0.3145 & 13.84 & 0.3595\\
		Ours & 0.0211& 0.2084& 17.59& 0.4951 \\
		\hline
	\end{tabular}
	\label{tab:2s}
\end{table}

 To further quantify the analysis, Table \ref{tab:2s} presents the average MSE, LPIPS, PSNR, and SSIM scores of each algorithm on the simulated test dataset with added camera noise. As shown in the table, with the introduction of camera noise, traditional methods like ADMM show significant deterioration across all metrics, resulting in a noticeable drop in image quality. Although FlatNet and DeepLIR make some improvements in denoising, they still fail to effectively restore details and color, leading to a decline in performance. MWDN demonstrates relatively stable performance in noise handling, but still struggles with fine detail recovery and image brightness enhancement. In contrast, the proposed method shows minimal degradation compared to the noise-free case, with particularly strong results in PSNR and LPIPS.

These results not only confirm the effectiveness of the proposed method in handling camera noise under low-light conditions, but also highlight the limitations of previous methods under the same conditions, further emphasizing the unique advantages of the proposed approach in solving image reconstruction under low-light environments.

\subsection{Measured Reconstruction}

\begin{figure}[t]
	\centering\includegraphics[width=10cm]{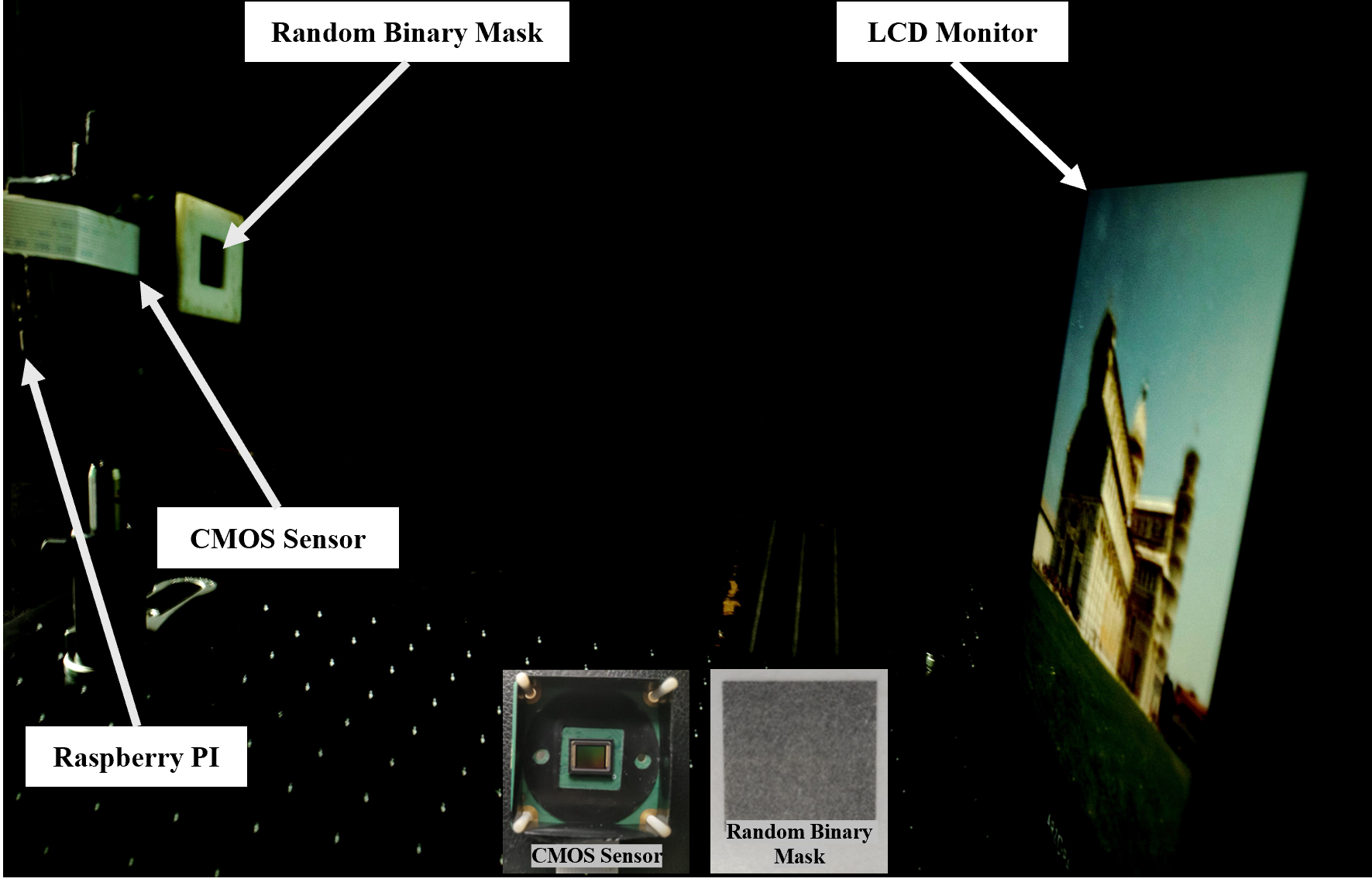}
	\caption{Our self-built lensless imaging system.}
	\label{f5}
\end{figure}

This section validates the proposed method through measured experiments. As shown in Fig. \ref{f5}, we placed a self-designed random binary mask in front of the CMOS sensor, with a distance of 10mm between the mask and the CMOS sensor, considering the thickness of the glass covering the CMOS surface and the mask. Then, an LCD display screen used to display the captured target images was positioned 300mm in front of the CMOS sensor. The distance between the screen and the sensor is optimal for our imaging device. The system has a field of view (FOV) of about 26.6°. The target scene is placed at a distance that matches the adopted PSF, ensuring optimal imaging. If the scene is positioned outside this range, image quality degrades due to a mismatch between the assumed and actual system response. The raw resolution collected by the CMOS sensor is 4056×3040, encompassing Bayer measurements with four original channels (R, Gr, B, Gb), and the Bayer array image was converted to an RGB image with the dimensions of 2028 × 1520 × 3.

The specific acquisition process for the measured data involves configuring the exposure time and exposure of the CMOS sensor using a Raspberry Pi to 0.7s and 100, respectively. The CMOS sensor is then set to collect data every 10 seconds, and the collected raw measurement data is saved to a computer. The captured target images are switched on the LCD display screen every 10 seconds until all target images in the original dataset have been traversed, resulting in the final dataset for the measured experiments.

\begin{figure}[t]
	\centering\includegraphics[width=\textwidth]{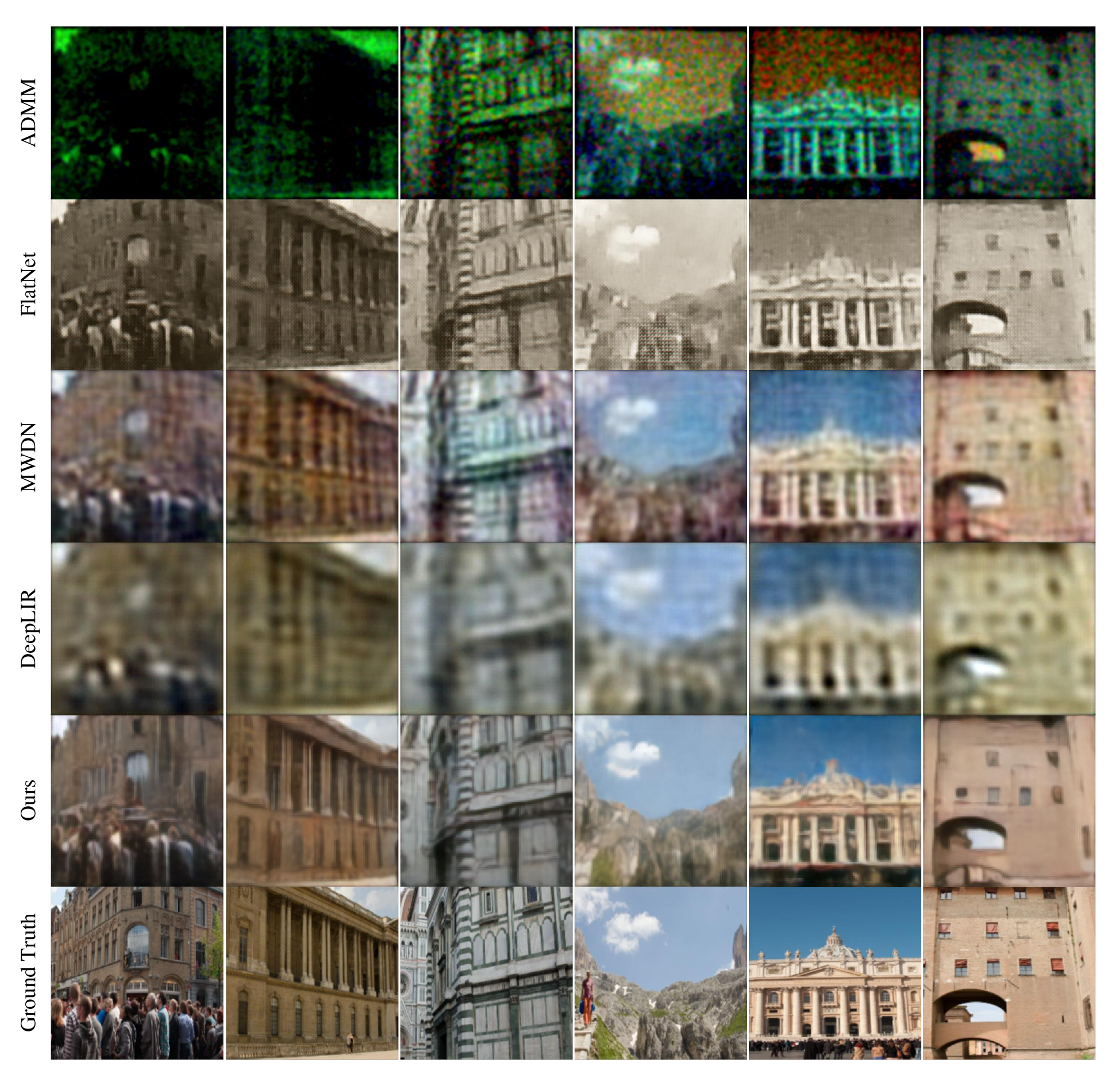}
	\caption{ The test set results of the self-built measured dataset, from top to bottom, are the ADMM, FlatNet, MWDN, DeepLIR, Ours, and Ground Truth.}
	\label{f6}
\end{figure}

\begin{table}[htbp]
	\centering
	\caption{\bf The average MSE, LPIPS, PSNR and SSIM of the proposed method and several other methods on the simulation test set.}
	\begin{tabular}{ccccc}
		\hline
		Method  & MSE & LPIPS & PSNR(in dB) & SSIM\\
		\hline
		ADMM    & 0.1371& 0.5710 & 8.76 & 0.1952\\
		FlatNet & 0.0180 & 0.1646 & 18.35 & 0.4952\\
        MWDN   & 0.0118 & 0.1965 & 19.56 & 0.5630\\
        DeepLIR  & 0.0126 & 0.1885 & 19.21 & 0.5166\\
		Ours & 0.0071& 0.1325& 22.02& 0.6392 \\
		\hline
	\end{tabular}
	\label{tab:3s}
\end{table}

\begin{figure}[t]
	\centering\includegraphics[width=\textwidth]{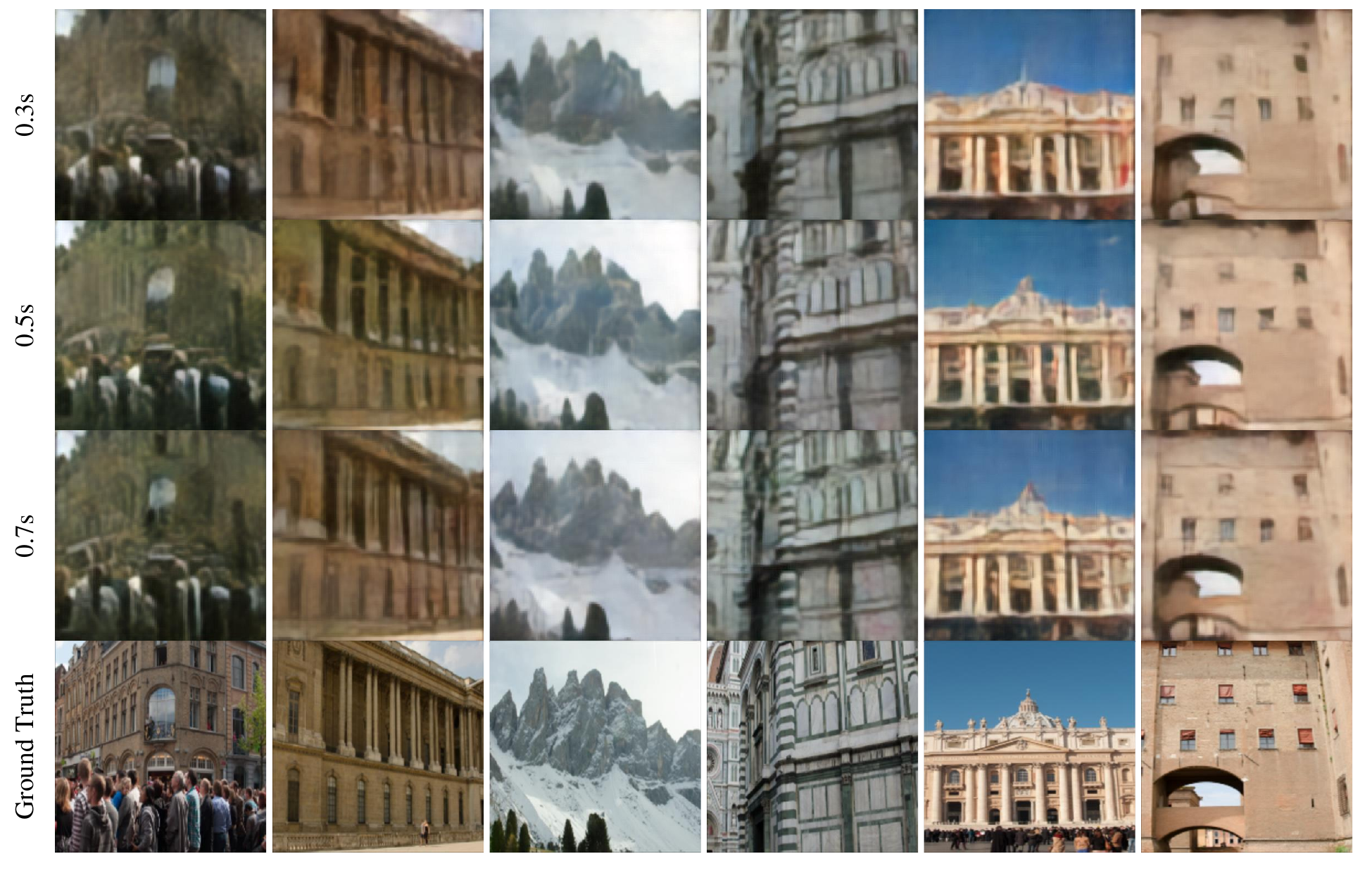}
	\caption{Reconstruct and enhance results for measured datasets acquired at varying exposure, from top to bottom, are the results for the 0.3s exposure, the 0.5s exposure, the 0.7s exposure, and the ground truth.}
	\label{f7}
\end{figure}

To evaluate the performance of the proposed reconstruction enhancement algorithm, experiments were conducted on the real-world test set, using ADMM with 100 iterations, FlatNet, MWDN, and DeepLIR  as baseline methods. Fig.~\ref{f6} presents some sample reconstruction results from the real-world validation dataset, along with visual comparisons to the original input and ground truth images.

As shown in this figure, ADMM produces poor reconstruction quality, only recovering the basic contours of the target, with the image almost entirely overwhelmed by noise. FlatNet and DeepLIR, while effectively removing noticeable noise, suffer from significant loss of color and detail information, resulting in subpar reconstructions. MWDN performs relatively well in preserving color information and recovering the image to some extent, but still struggles with fine details. In contrast, the proposed algorithm significantly improves reconstruction quality, virtually eliminating noise, and effectively preserving both color and detail information, yielding superior visual results compared to the other methods.

To further quantify the analysis, Table \ref{tab:3s} presents the average MSE, LPIPS, PSNR, and SSIM scores for each algorithm on the real-world test dataset. As shown in the table, ADMM performs the worst across all metrics, with very low scores. In comparison, the two-stage networks FlatNet and DeepLIR show significant improvements over ADMM, but still slightly lag behind MWDN. MWDN achieves the best performance among the baseline methods, particularly showing a notable improvement in SSIM. In contrast, the proposed method outperforms all other methods across all metrics, consistent with the visual results in Fig.~\ref{f6}, demonstrating its clear advantage in reconstruction quality.

The results on the measured test set verify the excellent effectiveness of the proposed method in low-light conditions is also verified, and the limitations and shortcomings of previous imaging methods under the same conditions are profoundly revealed, and the unique advantages of the proposed method in solving the problem of image reconstruction in low-light conditions are further highlighted through the comparative analyses.

To evaluate the robustness of the proposed reconstruction enhancement algorithm under varying low-light conditions, three sets of measured data were collected with exposure times of 0.7s, 0.5s, and 0.3s, respectively. Fig.~\ref{f7} displays sample reconstruction results from these datasets. Visually, the reconstructed images exhibit no significant differences in detail or color information across the three exposure conditions. All results achieve satisfactory reconstruction and enhancement, indicating that the algorithm maintains high robustness under different low-light scenarios.

For a more detailed evaluation, the average MSE, LPIPS, PSNR, and SSIM metrics of the reconstructed images under each exposure condition were calculated and are presented in Table \ref{table:4s}. As the exposure time decreases from 0.7s to 0.3s, these metrics show only minor declines, with no significant performance degradation. This is consistent with the visual results in Fig.~\ref{f7}, confirming that the proposed algorithm effectively preserves image quality even under reduced exposure conditions. These findings underscore the excellent performance and stability of the method when applied to varying low-light environments.

\begin{table}[htbp]
	\centering
	\caption{\bf The average MSE, LPIPS, PSNR and SSIM of the proposed method on the simulation test set under different low light conditions.}
	\begin{tabular}{ccccc}
		\hline
		exposure times(s)  & MSE & LPIPS & PSNR(in dB) & SSIM\\
		\hline
		0.3    &  0.0078& 0.1417 & 21.55 & 0.6271\\
		0.5 & 0.0073 & 0.1373 & 21.85 & 0.6365\\
		0.7 & 0.0071& 0.1325& 22.02& 0.6392 \\
		\hline
	\end{tabular}
	\label{table:4s}
\end{table}

On the dataset used in this study, the proposed method takes approximately 0.4s to reconstruct a single target image, with a memory usage of around 4GB. This demonstrates that the method strikes a balance between performance and computational efficiency, making it suitable for practical applications. Furthermore, the efficiency of the method during the diffusion process further underscores its applicability in large-scale data processing scenarios.

\section{Discussion}

	In this work, we focus on addressing the challenges of lensless imaging under low-light conditions, with an emphasis on improving image reconstruction methods. Current lensless imaging techniques predominantly rely on coded-aperture light modulation, which can be broadly categorized into amplitude masks and phase masks. Phase masks, which modulate the phase of incident light instead of its amplitude, generally offer higher light throughput. This characteristic makes them more suitable for low-light scenarios compared to amplitude masks. However, despite their advantages, phase-mask-based systems often fail to match the reconstruction quality of traditional lens cameras, necessitating further advancements in reconstruction algorithms.

\begin{figure}[t]
	\centering\includegraphics[width=\textwidth]{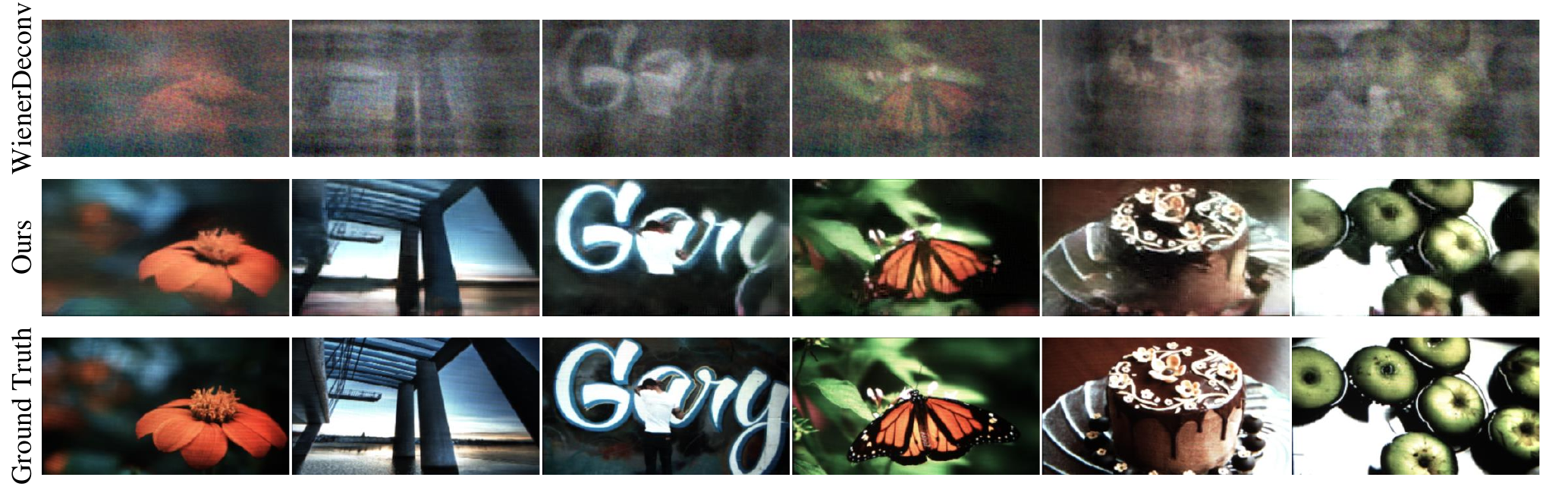}
	\caption{Reconstruction results on the diffusercam dataset with added camera noise in low-light conditions.}
	\label{f8}
\end{figure}

	To evaluate the proposed method, we conducted experiments using a self-built amplitude-mask-based lensless camera. Additionally, to assess its performance on phase-mask systems, we utilized the publicly available DiffuserCam dataset \cite{monakhova2019learned}. While the original dataset was captured under normal lighting, we simulated low-light conditions using a camera noise model described in Section \ref{s0}. Experimental results, illustrated in Fig.~\ref{f8}, compare our method with traditional Wiener deconvolution. Under low-light conditions, phase-mask systems exhibit significant noise and insufficient brightness. In contrast, our method improves image brightness, effectively suppresses noise, and produces visually realistic reconstructions, as demonstrated in the second row of Fig.~\ref{f8}.

	Low-light imaging presents challenges that extend beyond the mere issue of insufficient illumination. In such scenarios, the interaction between faint target light and varying environmental light conditions can introduce additional complexities, such as uneven illumination, color distortion, and interference. Most lensless imaging systems, including ours, are typically tested in controlled indoor environments, where light conditions can be precisely managed. These experiments often involve re-photographing scenes displayed on monitors to minimize environmental light interference.

	A recent study addressed this issue \cite{chakravarthula2023thin}, sharing a similar conceptual framework with us, by employing a diffusion model conditioned on imaging results influenced by outdoor environmental lighting. In  addition, they augmented their system with an array of metalenses to gather additional information, enabling promising results under real-world broadband illumination. Their work also underscores the importance of multiplexed measurements, integrating hardware enhancements, such as custom-designed nanophotonic arrays in their way, to modulate supplementary information for computational imaging.

	Future low-light applications of lensless imaging outside laboratory settings face dual challenges: insufficient target light and interference from environmental light. These issues suggest that multiplexed measurements could play a critical role in overcoming these limitations. However, implementing such measurements in physical systems remains an open problem, particularly given the spatial constraints inherent in low-light applications. Addressing these challenges will require innovative approaches to optimize both hardware design and computational algorithms, paving the way for robust lensless imaging systems suitable for real-world environments.

\section{Conclusion}
In summary, this paper proposed an innovative two-stage, model-driven generative reconstruction framework for lensless high-quality reconstruction under low-light conditions. In the first stage, a learnable Wiener filter-based module generates an initial, noisy reconstruction. The result is then transformed into the wavelet domain using a 2D discrete wavelet transform, producing lower-dimensional subbands for efficient processing. In the second stage, a noise-robust conditional diffusion generative model is applied to progressively refine the reconstruction, incorporating forward diffusion and backward denoising during training to ensure stable outputs. The experimental results show that the proposed method provides a substantial improvement in image brightness, noise reduction and overall sharpness in low-light conditions. It also reveals the limitations in previous reconstruction approaches, and demonstrates the unique advantages of the proposed method in solving the image reconstruction problem in low-light conditions.

\begin{backmatter}
	\bmsection{Funding}This work was supported in part by the National Natural Science Foundation of China under Grants 62471113, 62305049 and 62371104, and in part by Sichuan Science and Technology Program 2024NS-FSC0479 and 2024NS-FSC1439.
	
	\bmsection{Disclosures}The authors declare that there are no conflicts of interest.
	
	\bmsection{Data availability}Data underlying the results presented in this paper are not publicly available at this time but may be obtained from the authors upon reasonable request.
\end{backmatter}

\bibliography{manuscript}
\end{document}